# *Ultra-compact nonvolatile plasmonic phase change modulators and switches with dual electrical-optical functionality*

Jacek Gosciniak

*Independent Researcher, 90-132 Lodz, Poland*
*Email: jeckug10@yahoo.com.sg*

**Abstract**
Programmable photonic integrated circuits (PICs) are the foundation of on-chip optical technologies with the optical modulators being one of the main building blocks of such programmable PICs. However, most of the available modulators suffer from high power consumption, low response time and large footprint. Additionally, they show a large resistance modulation, thus they require high switching voltage. In consequence, they operate much above CMOS-compatible voltages of 1.2 V and with high insertion losses. Furthermore, the state and information they carry are lost once the power is turned off – so, they are volatile. Thus, realizing modulators and phase shifters that overcome all of those problems still remain a challenge.

To overcome some of those limitations the nonvolatile phase change materials implemented in the plasmonic structures are proposed that can offer many advantages as a result of high electric field interaction with nonvolatile materials. Consequently, proposed here novel plasmonic nonvolatile switches can operate by phase modulation, absorption modulation, or both and under zero-static power. Thus, only 230 nm long active waveguide is needed to attain full $\pi$ phase delay with an insertion loss of 0.12 dB. Apart from it, when operating by amplitude modulation an extinction ration exceeding 2.2 dB/µm can be achieved while an insertion loss is kept at 0.185 dB/µm. Furthermore, the heating mechanism can be based on the external heaters, internal heaters, electrical (memory) switching or optical switching mechanism what provide a lot of flexibility in terms of a design and requirements.

**Introduction**
In the last years, photonic integrated circuits (PICs) became very attractive as they offer broad bandwidth and very efficient information transport, processing, and storage [1, 2]. Large-scale PICs are based on the silicon photonics platform that is compatible with the well-established CMOS technology. To realize large-scale photonic systems, active components such as photodetectors, modulators and switches are very essential to detect and control the light flow within the network [3]. Apart from the photodetectors that performs optical to electrical signal conversion, the on-chip modulators and switches convert the electrical signals into the optical ones, and thus they are key components in photonic links. They should be characterized by low static and dynamic power consumption, high switching contrast, low optical loss, compact footprint, dense integration and ultrafast switching speed. Furthermore, they should operate under CMOS driving voltages of 1.2 V.

Most of the available modulators and switches can be classified as volatile devices as they require a constant supply of the electric power to hold each switching state. However, for high-performance operations and neuromorphic computing, the nonvolatile functionalities are essential. A term nonvolatile means that no static energy or holding power is required to retain any of the states once it is set. In consequence, it enables a low power consumption as no power is required to keep a switching state and brings new possibilities to the existing platform [4]. Huge impact is expected to technologies such as programmable photonics [5], photonic memories [6], neural networks [7], LIDAR systems [8], or power-efficient switching in data centers [9].

In terms of a switching mechanism, modulators can operate based on the plasma dispersion effect [1, 10], thermo-optic (TO) effect [11, 12, 13, 14], or electro-optic (EO) effect [15, 16] where the modulation is achieved by either changing the real part (*n*) of the modal refractive index leading to optical modulators (OM) [16, 17], or by modulating the imaginary part (*k*) of the modal refractive index leading to absorption modulators (AM) [18]. Thus, OMs relate to the phase of the light, whereas AMs to the intensity absorption of the light. In both types, the fundamental complex index of refraction is altered thermally or electrically in the active material, which in turn modifies the propagation constant of the mode inside the respective waveguide.

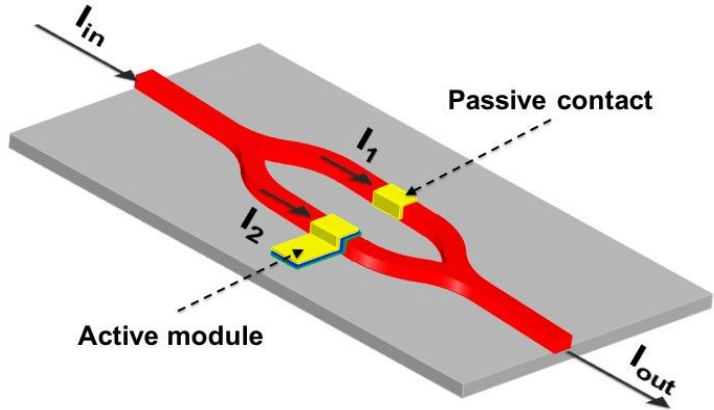

**Figure 1.** Perspective view of the Mach-Zehnder interferometric modulator with the active biasing contacts and passive metallic contact for the field loss balancing.

For the optimal operation conditions, the phase modulators require a large variation of the real part of the mode effective index ($\Delta n$) while losses should stay low, *i.e.*, small variation of the imaginary part between on and off states ($\Delta k = 0$). On the other hand, the intensity modulators require a large variation of the imaginary part ($\Delta k$) with the minimum losses, *i.e.*, minimum *k* for on or off states. Furthermore, the phase modulation always requires a reference phase for comparison, thus the modulator requires the interferometric arrangement [12, 13, 17, 19, 20] (**Fig. 1**). In comparison, absorption modulators can operate under even linear waveguide arrangement.

Thermo-optic Mach-Zehnder interferometers (MZIs) or micro-ring resonators (MRRs) are the most extensively studied TO phase modulators (switches) where the index of propagating modes is changed through resistive metal contact on top of waveguides [19]. While MZI are relatively temperature insensitive and provides very high bandwidth of modulation, they suffer however from high power switching, low switching time and large footprint [21]. In contrary, MRRs can provide lower power consumption, reduced footprint and still very high bandwidth, however they are not as temperature insensitive as MZI and require more fabrication precision [21].

In electro-optic modulators the electrical signal modifies the refractive index of the material through the Pockels effect, Kerr effect, quantum confined Stark effect, or free carrier dispersion effect [16]. In terms of the modulators and switches that are based on free carrier dispersion effect, they show low power consumption and faster modulation speed, however, they suffer from large device footprint as a result of small refractive index change ($< 10^{-3}$) [3]. Furthermore, each modulation schema requires specific application-defined design.

Thus, most modulators and switches suffer either from a high power consumption [19], a low operation speed [11, 12, 13], a high optical insertion loss [17], a low modulation depth, or a large footprint [20]. Furthermore, some of them are not suitable for operation at temperature above 100 °C due to a material degradation tendency at moderate temperatures [22]. Finally, all of them require a constant supply of the electric power as they are volatile. As a result, finding a better material and design platform is of great interest for the industry. In the last few decades, it has become obvious that further progress can be made through an implementation of new materials [23, 24, 25, 26, 28], integration with plasmonics [29, 30], or both.

## Plasmonics platform: LR-DLSPP arrangement

Plasmonics can squeeze light much below the diffraction limit, which reduces the device footprint [29]. Furthermore, a small device volume means a higher density of integration and, simultaneously, lower power consumption, easier heat dissipation, and faster operation speed [30].

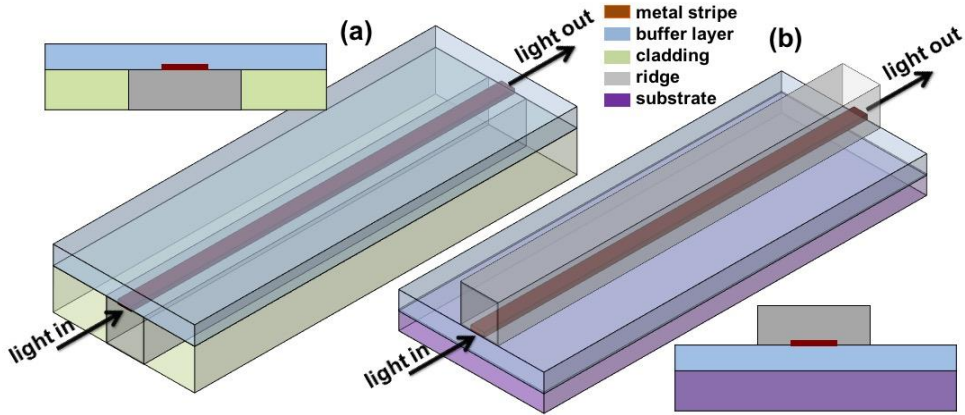

**Figure 2.** Proposed LR-DLSPP waveguide configuration in (a) "inverse" and (b) "normal" designs.

The proposed modulators are based on the long-range dielectric-loaded surface plasmon polariton (LR-DLSPP) waveguide arrangement with the metal stripe placed between the ridge and buffer layer that support TM-polarized mode (**Fig. 2**). Thus, the metal stripe is an essential part of the LR-DLSPP waveguide and can serve as one of the electrodes [31, 32, 33, 34].

In a balance conditions, the mode effective index below a metal stripe is close to the mode effective index above a metal stripe. In consequence, the absorption in metal stripe is minimalized and the propagation length is enhanced. When the active material is deposited either on top of the buffer layer in "inverse" design (**Fig. 2a**) or on top of the ridge in "normal" design (**Fig. 2b**) even the small change in the refractive index of the active material can disturb a balance (**Fig. 3**). Depending on the refractive index of a buffer layer and a ridge, the mode is pushed either to the ridge or the buffer layer (**Fig. 3a**). As a result, the absorption in metal arises and propagation length of the mode decreases. This effect can be highly enhanced when the active material is placed directly at the contact with a metal stripe, *i.e.*, in the electric field maximum of the propagating mode.

The simulations were performed for the ridge width of 300 nm and heights of 140 nm and 180 nm while the buffer layer thickness was kept constant at 120 nm. As all parameters were chosen rather randomly, there is plenty of room to optimize a design. Two different scenarios (cases) of PCM location were considered here – in a first case the PCM constituted for the buffer layer (**Fig. 3a**) and in the second case it constituted for a ridge (**Fig. 3b**). As the second material that constitutes for a LR-DLSPP waveguide, the Si was chosen due to a refractive index that is close to most of PCMs. To avoid oxidation from air and, consequently, excessive optical losses, the PCM was covered by $SiO_2$ that covered additionally the Si.

The proposed LR-DLSPP waveguide ensures low attenuation below 0.0075 dB/μm that can be achieved with Si platform (**Fig. 3a**). Consequently, assuming even 10 μm long active region of the modulator, the insertion losses (*IL*) below 0.075 dB can be achieved. Even farther reduction in absorption losses can be achieved with lower index waveguide materials where the propagation length of 700 μm was measured at telecom wavelength for CMOS-compatible silicon nitride (SiN) [25].

Furthermore, the proposed plasmonic waveguide ensures high coupling efficiency with the photonic platform that was numerically estimated at 97 % (**Fig. 3a**) [31, 32]. Thus, the coupling losses per interface as low as 0.05 dB can be achieved. It was experimentally validated, where the coupling efficiency exceeding 75 % per interface was achieved. A difference between numerical calculations and experimental results was attributed to the presents of thin layer of titanium used for improving adhesion between gold stripe and substrate that introduces substantial mode absorption [34].

In consequence, apart from an efficient conversion mechanism, the proposed plasmonic waveguide provides an excellent platform for a realization of on-chip modulators and switches.

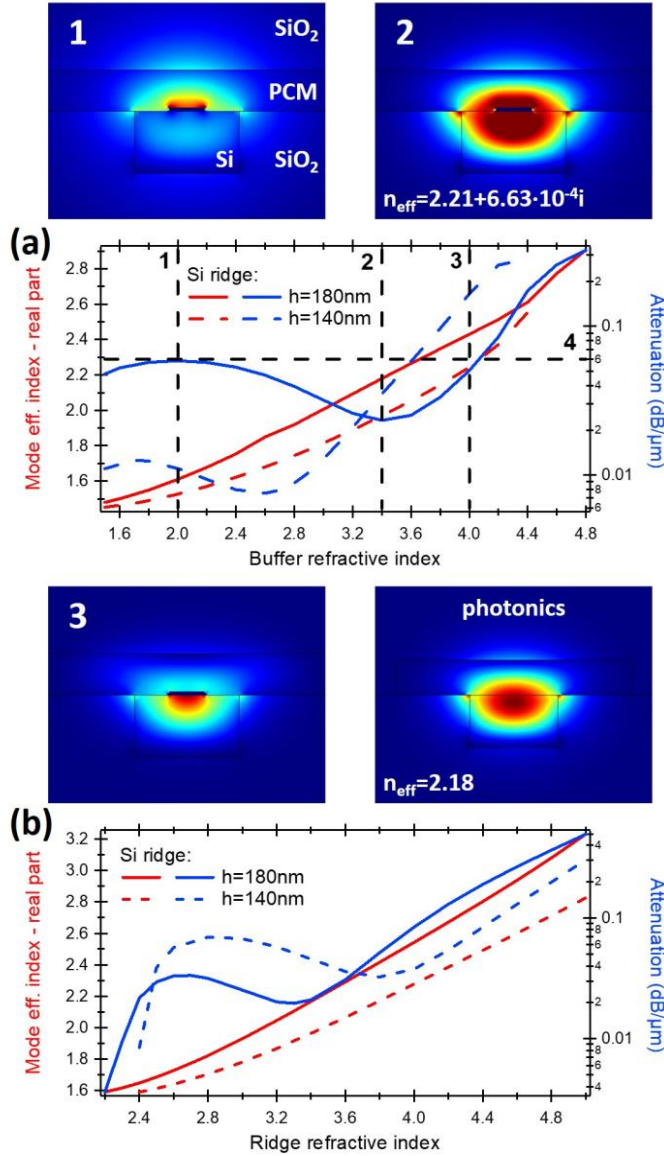

**Figure 3.** (a, b) Mode effective index (red lines) and attenuation (blue lines) as a function of (a) buffer layer refractive index and (b) ridge refractive index calculated for LR-DLSPP waveguide with Au stripe. The Si ridge dimensions were taken at $w = 300$ nm, $h = 140$ nm and $180$ nm while the buffer layer width and thickness were kept at $w = 800$ nm and $h = 120$ nm. The Au metal stripe dimensions were taken at $w = 100$ nm, $h = 10$ nm.

**Material platform: phase change materials**

Compared to the volatile modulators and switches that requires a constant power supply to maintain the switching state, the nonvolatile optical modulators and switches no necessitate static energy or holding power (continuous power supply) to retain (keep) the switching state. Recently, the phase change materials (PCMs) have been proposed as very promising materials for realization of nonvolatile optical modulators and switches [23, 26, 35, 36].

They provide extremely high refractive index contrast ($\Delta n = 0.6 \div 3.0$), ultrafast transition (> 1 ns), energy-efficient reversible switching, nonvolatility (leading to zero-static power consumption), high scalability, long-term retention (< 10 years), and high cyclability ($10^{12}$ switching cycles) [37, 38, 39, 40]. Mechanism of switching in the PCM is realized through a phase transition between its two different phases – amorphous and crystalline [38, 39]. It can be performed by Joule heating of the PCM using external heaters (thermal effect) [41, 42], electrical pulses (electro-thermal effect) [23, 30, 43], or optical pulses (photo-thermal effect) [26, 27, 30, 36]. In photonic nonvolatile phase-change switches, the heat-induced refractive index change of PCM induces change in transmitted light.

$Ge_2Sb_2Te_5$ (GST) and $Ge_2Sb_2Se_4Te_1$ (GSST) are the most commonly used PCMs for integrated phase-change devices owing to its wide availability, well established technology, relatively low switching temperature and very large refractive index shift [38, 42, 43]. The crystalline state in PCMs is achieved by heating the material above the crystallization temperature of 160 °C while the amorphization is achieved by heating the material above the melting temperature of 600 °C and then quenching by rapid cooling (nanoseconds) [4]. In terms of switching time between states, it is generally slower for crystallization and physically limited to hundreds of picoseconds [44].

The refractive index contrast between an amorphous and crystalline states at wavelength of 1550 nm for GST was calculated at $\Delta n = \sim 2.74$ while for GSTT at $\Delta n = \sim 2.03$. As observed (**Fig. 4a**), the refractive index of both a-GST and a-GSST is very close to the refractive index of Si, $n = 3.47$ what provides a good mode matching. However, both materials show large absorption losses in the crystalline phase with $k = \sim 1.09$ and $k = \sim 0.42$ for c-GST and c-GSST, respectively. As a result, a transmitted light through GST- or GSST-based waveguides undergo much larger change in amplitude than in phase during a transition. This limits their use to amplitude modulation rather than to phase modulation.

Recently, a new class of low loss PCMs such as, for example $Sb_2S_3$ and $Sb_2Se_3$ were demonstrated that can be considered as reversible alternatives to the standard commercially available chalcogenide GST and GSST [28]. They offer zero loss in both amorphous and crystalline states at both 1310 nm and 1550 nm and a large index change what makes it an ideal candidate for nonvolatile phase change switches (**Fig. 4b**). At a wavelength of 1550 nm, $Sb_2S_3$ shows a complex refractive index $n = 2.71+0i$ for amorphous phase (a-SbS) and $n = 3.31+0i$ for crystalline state (c-SbS) while for $Sb_2Se_3$ it is $n = 3.28+0i$ and $n = 4.05+0i$ for amorphous (a-SbSe) and crystalline (c-SbSe) phases, respectively. Thus, a contrast of refractive index of $\Delta n = 0.60$ for $Sb_2S_3$ and $\Delta n = 0.77$ for $Sb_2Se_3$ is achieved while maintaining very low losses, $k < 10^{-5}$ (**Fig. 4b**).

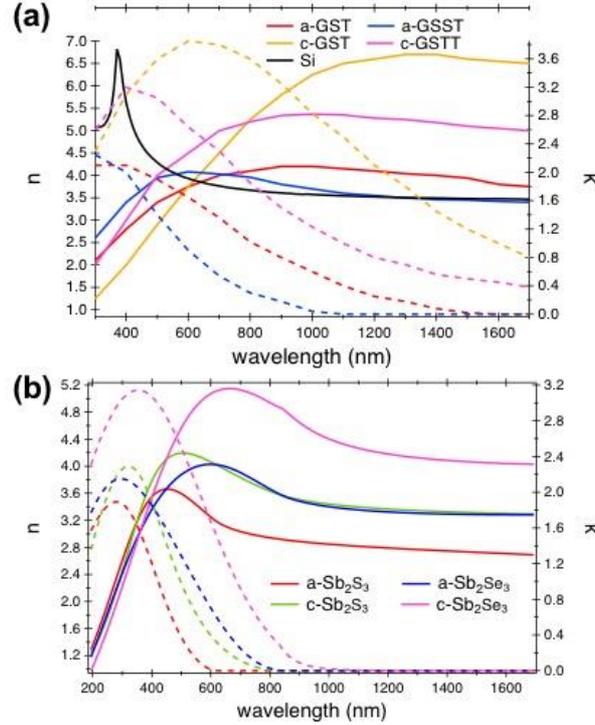

**Figure 4.** Real (solid lines) and imaginary (dashed lines) parts of refractive index of the amorphous and crystalline phased of (a) $Ge_2Sb_2Te_5$ (GST), $Ge_2Sb_2Se_4Te_1$ (GSST) [26], and (b) $Sb_2S_3$, $Sb_2Se_3$ [28] from the visible range to near-infrared. The results were compared with Si.

**Modulator arrangement**

In photonic nonvolatile PCMs arrangement the complex optical refractive index is tuned which affects both the optical absorption level and the phase and, therefore the transmitted light.

The proposed PCM-based plasmonic modulator enables reversible switching the state of PCM between its amorphous (high resistance, high transmission) and crystalline (low resistance, low transmission) states by sending either electrical pulses through the metal electrodes or optical pulses through the waveguide. When electrical current flows through PCM that forms the circuit, the Joule heating of PCM causes the change of the phase. This switching method is often called "memory switching". However, the main limitation with this type of switching mechanism is that only a small volume of the PCM can be switched due to a filamentation [45]. To switch larger area of PCM, the state of PCM can be switched by Joule heating using one of the metal electrodes that can work as a heater.

In most of the presented up to day PCM switches the heating mechanism need quite some extra space, which reduces computing density (TMACs/s/mm$^2$) and also increases insertion loss. Thus, the approach of using PCM on integrated photonics still needs to understand how to do that in an ultra-compact and scalable way.

The metal stripe is an essential part of the proposed LR-DLSPP waveguide and can be implemented as an internal heater [31, 32, 33, 34]. As it is in direct contact with a PCM, the heating process can be very efficient. Furthermore, the metal stripe is placed in the electric field maximum of the propagating mode, thus even small change in a refractive index of PCM close to the metal stripe can highly influence the propagating mode. However, to enhance a performance of the PCM-based modulator a few aspects of the heat flow should be considered. Under an applied voltage to the metal stripe, the electrical energy is dissipated into heat. The heat from a metal stripe dissipates to any materials that are in contact with the metal through conductive heat transfer [14]. The amount of heat transfer to the area of interest (ridge or buffer layer) depends upon the thermal conductivity coefficients of the materials that are in contact with metal stripe, contact area and thickness of the buffer layer and ridge. Thus, to ensure an efficient heat transfer to the PCM the second material that constitutes for LR-DLSPP waveguide should possess lower thermal conductivity coefficient [14].

The proposed LR-DLSPP-based modulator can be arranged as well with the external heaters as shown in **Fig. 5 and 6**. They can be arranged in two different configurations: lateral (**Fig. 5a and 6b**) and vertical (**Fig. 5b and 6a**). The lateral configuration consists of two resistive heaters placed directly in contact with the PCM (**Fig. 5a and 6b**) that is part of the waveguide, thus providing more heat to the PCM locally what lowers the switching threshold. In comparison, the vertical configuration consists of metal heating electrode placed on top of the either buffer layer (**Fig. 5b**) or ridge (**Fig. 6a**). Depending on the requirements, the heating electrode can be separated from PCM layer through very thin conductive layer to minimize the mode-overlap with the metal and eventual scattering while still providing an efficient heat transfer to the PCM.

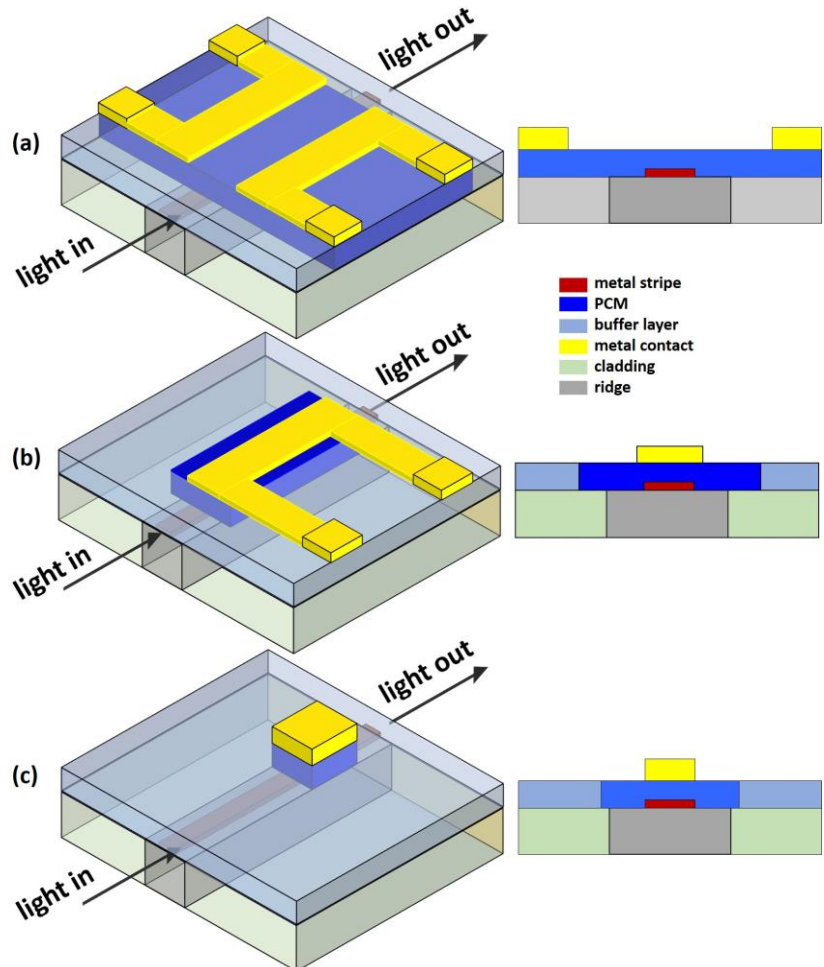

**Figure 5.** The active region of PCM-based modulators in „inverse" design with (a) lateral and (b, c) vertical electrodes arrangements.

As the LR-DLSPP mode is tightly bounded to the metal stripe, the external electrodes can be placed very close to the propagating mode without influencing the propagation length. Thus, the ohmic losses due to the presence of external metals are minimalized. As it has been previously shown for a lateral configuration [46], the external electrodes can be placed as close as 50 nm away from the ridge without introducing any additional losses. For comparison, an essential increase of additional losses was observed for a similar arrangement but realized with Si photonic waveguide were presence of external electrodes contributed to 0.11 dB/µm additional losses for propagating TM mode [42].

Direct contact of PCM with the electrode allows lowering the threshold voltage for delivering the right amount of heat for inducing a phase transition in the PCM. A resistive heater optimized for efficient switching and contemporary not generating insertion losses can be made in doped silicon or in silicide positioned next to the waveguide [35]. Furthermore, the external heater can be made from transparent conductive oxides (TCOs) (**Fig. 5b and 6a**) such as for example ITO [41, 42] that is characterized by low optical losses at 1550 nm [17, 47]. Apart from it, graphene can be very efficient platform for realization of such a task [23, 24].

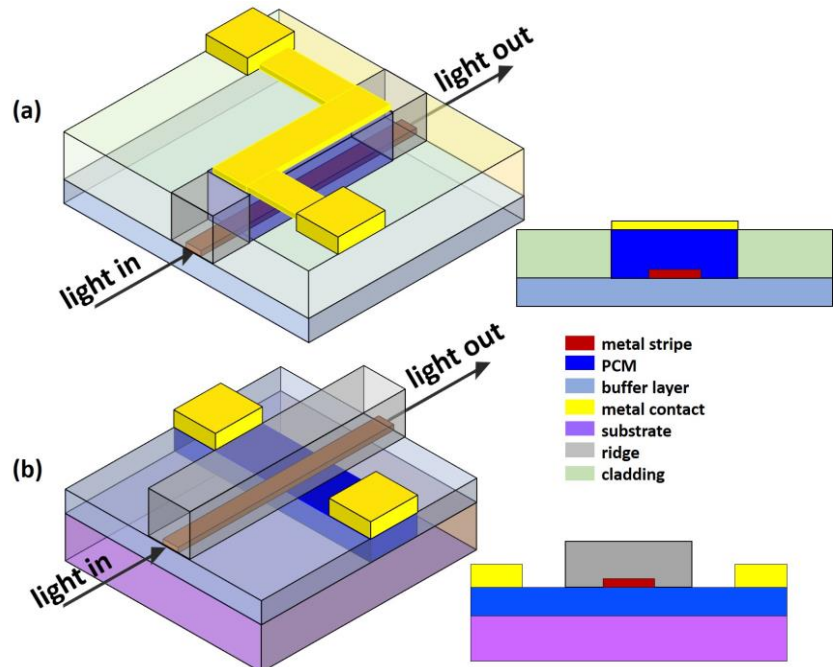

**Figure 6.** The active region of PCM-based modulator in „normal" design with (a) vertical and (b) lateral electrodes arrangements.

Moreover, the proposed arrangements provide strong light confinement inside the PCM and without introducing excessive parasitic optical losses. Furthermore, the heat is highly localized within PCM, thus the thermal mass is significantly lowered. In consequence, the switching voltage and current of the device can be highly minimalized.

**Results**
**Electrodes arrangement**
Under a heating of PCM through the light (optical pulses) or electrical contacts, the PCM undergoes a transition to a crystalline state that is characterized by much higher refractive index (**Table 1**). Thus, the balance in mode effective indices is broken as one side of the LR-DLSPP waveguide (upper or lower) with PCM shows higher mode effective index. In consequence, most of the optical energy is now pushed out of the PCM to the opposite side of the LR-DLSPP waveguide (**Fig. 3a**). As the balance is broken, the absorption losses in metal arises that give rise to the temperature increases of the metal stripe. The heat generated in such a way by a metal stripe dissipates its energy to any materials that are in direct contact with it. Depending on the materials surrounding the metal stripe, the amount of heat transferred to the PCM can change significantly. However, it provides an additional heat source that can provide further temperature increases of the PCM.

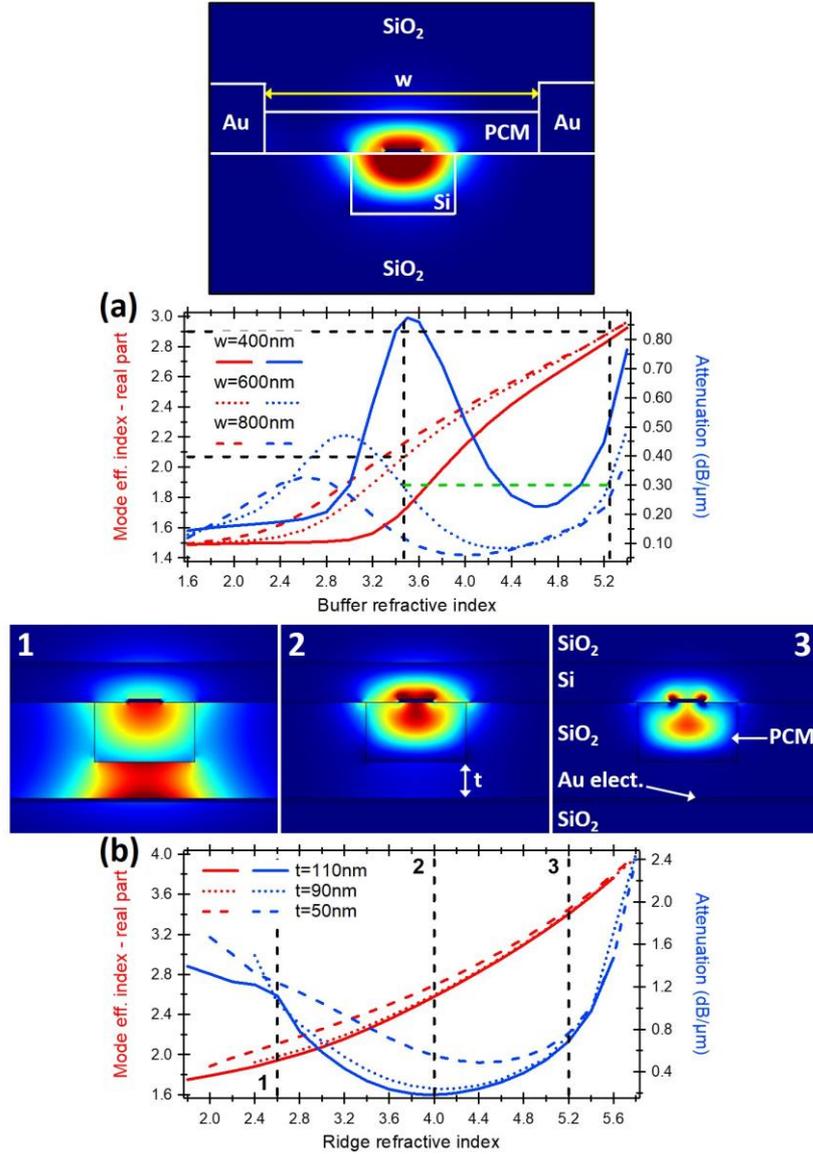

**Figure 7**. (a, b) Mode effective index (red lines) and attenuation (blue lines) as a function of (a) buffer layer refractive index and (b) ridge refractive index calculated for LR-DLSPP waveguide with external electrode(s). The Si ridge dimensions were taken at $w = 300$ nm, and $h = 180$ nm while the buffer layer width and thickness were kept at $w = 800$ nm and $h = 120$ nm. The Au metal stripe dimensions were taken at $w = 100$ nm, $h = 10$ nm. The calculations were performed for (a) different distance $w$ between external electrodes and (b) different distance $t$ between an external electrode and a ridge.

The proposed LR-DLSPP plasmonic mode is tightly bounded to the metal stripe, thus the external electrodes can be placed very close to the propagating mode without substantial influence on the attenuation (**Fig. 7**). In consequence, the ohmic losses due to the presence of external metals can be minimalized. As it is shown (**Fig. 7**), the external electrodes can be placed as close as 50 nm away from the ridge without introducing substantial losses even being 200 nm thick (**Fig. 7**) [46]. As such, in the lateral electrodes' arrangement (**Fig. 7a**), the minimum attenuation coefficient of 0.226 dB/μm was achieved for electrodes placed just only 50 nm away from the ridge. Further decreases of an attenuation was observed for longer electrodes spacing that drops below even 0.06 dB/μm for a distance between electrode and ridge at 250 nm.

Under a heating of PCM that constitutes of a buffer layer the refractive index of PCM increases. As observed from **Fig. 7a** for a distance between electrodes and the ridge of 150 nm, the change of a refractive index of the buffer layer from $n = 3.47$ to $n = 5.25$ corresponds to the same level of an attenuation of 0.3 dB/μm. However, the real part of the mode effective index in the same interval change from $n_{eff} = $ ~2.08 to $n_{eff} = $ ~2.91 that corresponds to the mode effective index change $\Delta n_{eff} = $ ~0.83 and a phase shift of ~1.07 π/μm. This corresponds to an ultra-compact phase shifter of only ~930 nm in length needed to attain full π phase delay. The small device footprint also yields a low theoretical insertion loss of 0.26 dB when switching the refractive index of PCM from $n = 3.47$ to $n = 5.25$. Even higher change of the mode effective index $\Delta n_{eff} = $ ~0.86 can be achieved for the electrodes placed 250 nm away from the ridge with a significant reduction of an attenuation that is kept constant at 0.19 dB/μm for a buffer later refractive index of $n = 3.2$ and $n = 5.0$. This change of mode effective index corresponds to a phase shift of ~1.11 π/μm and, consequently, requires only ~900 nm long phase shifter to provide a full π phase delay while the losses are kept at ~0.17 dB.

A distance of 50 nm between electrodes and the ridge is much smaller compared to a previously presented configuration with a Si rib waveguide, where a distance between electrodes and the rib was 500 nm [24]. The presence of metal electrodes contributed to an additional loss of 0.01 dB/μm while a presence of graphene that was used as a heater leaded to an additional loss of 0.1 dB/μm. Thus, the overall losses calculated for an amorphous state of GST exceeded 0.11 dB/μm for a distance between electrodes and a rib of 500 nm that is 10 time higher than in a proposed here LR-DLSSP waveguide organized in a lateral configuration. The insertion losses of ~0.11 dB/μm are only two times lower than ~0.226 dB/μm that was calculated for our configuration (**Fig. 7a**). Furthermore, a distance between electrodes and an active region of the waveguide influences the energy efficiency since part of the generated heat is dissipated around the long region. Thus, a shorter distance is highly desired as it enhances the heat transfer efficiency and reduces the power consumption. In Ref. 24 was stated that under a switching of GST from an amorphous to a crystalline state the real part of mode effective index changed from $n_{eff} = $ ~2.66 to $n_{eff} = $ ~2.87 while an attenuation arises from ~0.01 dB/μm to ~1.46 dB/μm. It corresponds to the mode effective index change $\Delta n_{eff} = $ ~0.21 and attenuation change of ~1.45 dB/μm. Thus, the presented configuration can serve only for an amplitude modulation.

In another example of the lateral electrodes configuration, a phosphorus-doped silicon was used as a heater [25]. Under a heating of $Sb_2Se_3$ and transition of phase from an amorphous to a crystalline, the mode effective index change $\Delta n_{eff} = $ ~0.071 was calculated that corresponds to a phase shift of ~0.09 π/μm. Thus, to attain full π phase shift a 11 μm long shifter is needed for which an insertion loss of ~0.1 dB was theoretically calculated. A small difference in attenuations between an amorphous and crystalline states of $Sb_2Se_3$ was calculated at ~0.007 dB for a 11 μm long shifter. However, a significant reduction in attenuations was possible as the metal contacts were far away from waveguide to reduce losses. It allows to avoid some additional losses but at the cost of power efficiency as more heat is dissipated before reaching an active part of the waveguide.

In comparison, the proposed in this paper arrangement with a lateral electrodes' configuration requires only ~900 nm long shifter to attain full π phase delay while losses for both states of PCM are kept constant at ~0.19 dB/μm. Thus, over 11 times reduction in length of a shifter was achieved while the losses are kept constant for both phases of the PCM. Furthermore, a direct contact of the external electrodes with the PCM and very close distance to the center of the waveguide provides huge benefits in terms of heat delivery and power consumption.

As mentioned earlier in the paper, the thickness of PCM for a proposed design can range from 50 nm to hundreds of nanometers that depends on where the PCM is deposited – in a buffer layer or a ridge. However, in most papers that can be find in a literature [24, 25, 42, 48, 49] the thickness of PCM is in a range of 10 - 30 nm as strong increase of an attenuation was observed for thicker PCMs [24]. For example, when $Sb_2S_3$ of different thickness was deposited on SOI waveguide the simulations performed for TE mode show the mode effective index change $\Delta n_{eff} = $ ~0.022 and change of attenuation from ~0.009 dB/μm to ~0.282 dB/μm for a 20 nm thick $Sb_2S_3$ and under a phase transition. However, for a 66 nm thick $Sb_2S_3$ on SOI waveguide the mode effective index change was calculated at $\Delta n_{eff} = $ ~0.038 while the attenuation grows from ~0.014 dB/μm for an amorphous phase to ~0.335 dB/μm for a crystalline phase. Even higher influence of the PCM thickness and width on the attenuation was observed for GST material where an increase of attenuation from 1 dB/μm for a 5 nm thick c-GST to up to 30 dB/μm for a 50 nm thick c-GST was calculated. In comparison, for a-GST the increase of attenuation was from 0.01 dB/μm to 0.1 dB/μm under increases of thickness from 5 nm to 50 nm [24]. Thus, to avoid additional losses most of the authors limit the thickness of PCM to 20 - 30 nm.

In a second proposed arrangement with a vertical electrode configuration and with the PCM placed in the ridge (**Fig. 7b**) an attenuation below ~0.185 dB/μm was calculated for a heating electrode placed only 110 nm away from the ridge and for PCM refractive index of $n = 4.0$. When the heating electrode is move closer to the ridge the attenuation arises and reaches ~0.48 dB/μm for an electrode placed just only 50 nm from the PCM ridge with a minimum corresponding to the PCM ridge refractive index of $n = 4.4$. As observed, the attenuation of ~0.43 dB/μm was calculated for a ridge refractive index of $n = 3.2$ and $n = 5$ that corresponds to the mode effective index change from $n_{eff} = $ ~2.15 to $n_{eff} = $ ~4.90. Thus, the change of the mode effective index exceeds $\Delta n_{eff} = $ ~2.75 and a phase shift ~3.55 π/μm. In consequence, only ~280 nm long active area is needed to attain full π phase delay with an insertion loss of ~0.12 dB.

For comparison, an essential increase of additional losses was observed for a similar arrangement but realized with Si photonic waveguide where a presence of external tungsten electrodes contributed to 0.11 dB/μm additional losses for a lateral arrangement and 0.06 dB/μm for a vertical arrangement with the heating electrode placed 250 nm above the silicon waveguide [42]. In this case, when a GSST transfer from an amorphous to a crystalline state under a heating approach, the mode effective index of the propagating TM mode changes from $n_{eff} = 1.942$ to $n_{eff} = 2.154+0.045i$. The change of the mode effective index can be calculated at $\Delta n = $ ~0.212 while an attenuation arises from 0 dB/μm to ~1.59 dB/μm. Thus, a proposed arrangement can be utilized to an absorption modulation rather than a phase modulation.

**Heat dissipation**
One of the ways to change the phase of the PCM can be realized through thermal heater. In such a case, the heating element that is properly biased dissipates energy in the form of Joule heat to the surrounding media.

**Table 1.** Thermal properties of the materials constituting the device. Here, $T_c$ is a crystallization temperature and $T_m$ is a melting temperature of PCM. The refractive indices were provided for wavelength of 1550nm.

| Material | Refractive index | Thermal cond. coeff. (W/m·K) | Heat capacity (J/g·K) | Density (g/cm$^3$) | $T_c$ (°C) | $T_m$ (°C) |
|---|---|---|---|---|---|---|
| Air | 1 | 0.026 | 1.005 | - | | |
| SiO$_2$ | 1.45 | 1.38 | 0.746 | 2.19 | | |
| Si | 3.47 | 148 | 0.72 | 2.32 | | |
| Si$_3$N$_4$ | 1.996 | 20 | 0.7 | 3.1 | | |
| Al$_2$O$_3$ | 1.65 | 30 | 0.9 | 3.9 | | |
| Au | 0.5958+10.92$i$ | 318 | 0.130 | 19.32 | | |
| ITO | 1.4+0.2$i$ | 3.2, 11, 1340 | 0.34 | 7.1 | | |
| a-GST | 3.80+0.025$i$ | 0.19 | 0.213 | 5.87 | 160 | 630 |
| c-GST | 6.63+1.089$i$ | 0.57 | 0.199 | 6.27 | | |
| a-GSST | 3.47+0.0002$i$ | 0.17 | 0.212 | 6.0 | 523 | 900 |
| c-GSST | 5.50+0.42$i$ | 0.43 | 0.212 | 6.15-6.3 | | |
| a-SbS | 2.71+0$i$ | - | - | - | 270 | 550 |
| c-SbS | 3.31+0$i$ | 1.16-1.2 | 0.353 | 4.6 | | |
| a-SbSe | 3.28+0$i$ | 0.36-1.9 | 0.507 | 5.81 | 180 | 620 |
| c-SbSe | 4.05+0$i$ | - | 0.574 | - | | |

Under an applied voltage to the metal stripe (**Fig. 8a**) or external electrodes (**Fig. 8b - e**), the electrical energy is dissipated into heat. The heat from metal electrodes (metal stripe or external electrodes) dissipate to any materials that are in contact with the metal through conductive heat transfer [14]. The amount of heat transfer to the area of interest (ridge or buffer layer) depends upon the thermal conductivity coefficients of any materials that are in contact with metal electrode, contact area and thickness of the materials. Thus, to ensure an efficient heat transfer to the PCM, the second material that is in contact with metal electrode should possess low thermal conductivity coefficient [14]. Furthermore, direct contact of PCM with the electrodes in the proposed modulator allows lowering the threshold voltage for delivering the right amount of heat for inducing a phase transition in the PCM.

To evaluate a heat transfer to the PCM the thermal resistance and thermal capacitance should be considered. The thermal resistance of an object is defined as $R_{th} = L/(\kappa \cdot A)$ and it describes the temperature difference that will cause the heat power of 1 Watt to flow between the object and its surroundings. In comparison, the thermal capacitance of an object, $C_{th} = C_p \cdot \rho \cdot V$, describes the energy required to change its temperature by 1 K, if no heat is exchanged with its surroundings [14]. Here, $L$ is the length of substrate along the heat transfer direction, $A$ is the cross-section area of the substrate, $V$ is the heated volume of the material, $\kappa$ is the thermal conductivity of the material, $C_p$ is the specific heat and $\rho$ is the mass density. Thus, the lower thermal capacitance the higher temperature rises for a given amount of heat delivered to the material. Furthermore, materials with lower thermal conductivity coefficient are characterized by higher thermal resistance, so lower electrical powers are required to increase a material temperature as the heat loss is reduced.

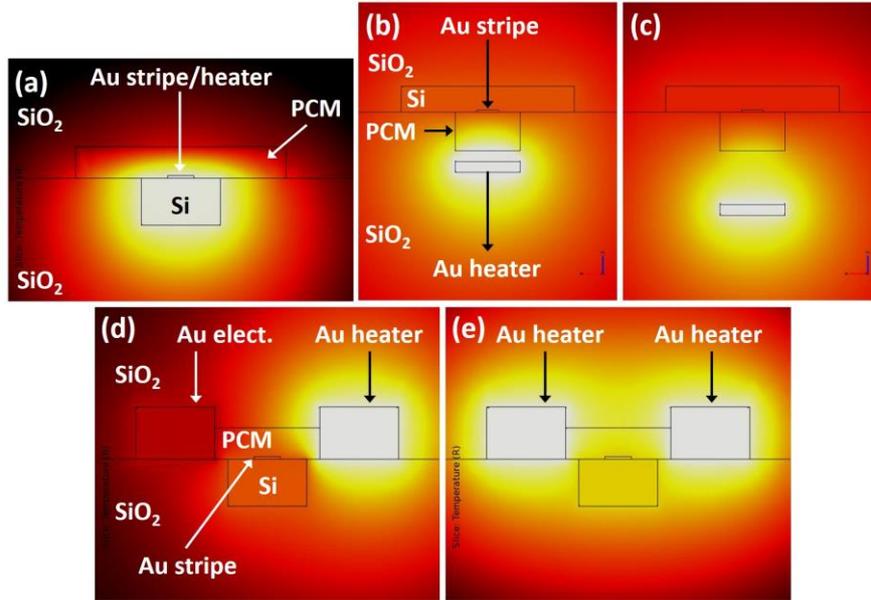

**Figure 8**. Heat map produced by Joule heating of the Au heating element in vertical (a, b, c) and lateral (d, e) configuration. In the electro-thermal simulation, the PCM is amorphous. For a vertical configuration the heating element was (a) part of the LR-DLSPP waveguide, (b) placed 50 nm away from the PCM ridge, (c) placed 250 nm away from a ridge. For a lateral configuration the PCM buffer was heated through (d) one heating element, (e) two heating element placed on both sides of a PCM buffer layer.

**Heating mechanism**
Here, the Joule heating process and heat dissipation model were performed using a 3D finite-element method simulation in COMSOL Multiphysics. We used the Joule Heating module where the $Sb_2Se_3$ were considered as the PCM. For a heating mechanism, three electrode configurations were considered: internal with a metal stripe (**Fig. 8a**), vertical (**Fig. 8b and c**) and lateral (**Fig. 8d and e**).

*a, External heater(s)*
The metal stripe that is a part of the LR-DLSPP waveguide can serve simultaneously as a heating electrode providing efficiently a heat to the area of interest – either to the ridge or a buffer layer (**Fig. 8a**). As it is placed in the electric field maximum of the propagating LR-DLSPP mode, even a small change of PCM refractive index will highly influence a propagating mode. Additionally, the heat is very efficiently provided to the PCM, thus high reduction in power consumption is expected. The amount of heat provided to the PCM can be highly enhanced by a proper choice of the material that is in direct contact with a metal stripe on the opposite interface. As observed from **Fig. 8a**, most of the heat from metal stipe is dissipated into Si ridge due to a huge difference in the thermal conductivities between PCM and Si. As observed from **Table 1**, the thermal conductivity coefficient of most PCMs range from 0.2 to 1.0 W m$^{-1}$ K$^{-1}$ while for Si reaches 148 W m$^{-1}$ K$^{-1}$. To enhance a heat transfer to the PCM a thin layer of material with a lower thermal conductivity coefficient compared to Si can be deposited between metal stripe and Si. However, this material should posse refractive index that is close to the refractive index of Si. In another solution, the Si can be replaced by SiN that is characterized by much lower thermal conductivity coefficient of 30 W m$^{-1}$ K$^{-1}$. As the refractive index of SiN is much lower compared to Si and most of the PCMs, the waveguide dimensions should be optimized.

In a case of the vertical configuration, the metal heating electrode is placed at certain distance from the PCM ridge that consists on the LR-DLSPP waveguide (**Fig. 8b, c**) and is surrounded by an oxide layer to avoid oxidation from the air and, consequently, an excessive optical losses. As showed in **Fig. 8b and c**, the heating electrode placed closer to the PCM ridge provide more efficient and faster heat transfer to the PCM. Thus, a lower power/voltage is required to increase a temperature of the PCM. This statement is pretty obvious, however in conventional photonic waveguides it is hard to realize it as the heating electrode have a strong effect on a propagating mode. When it is placed close to the photonic waveguide, the photonic mode is push to the electrode and, in consequence, the absorption losses highly arise. In comparison, the LR-DLSPP mode is strongly bounded to the metal stripe what allow to place an electrode close to LR-DLSPP waveguide without introducing significant losses (**Fig. 7a**).

In case of the lateral configuration, the heating electrodes are in direct contact with the PCM and are placed on both sides of the LR-DLSPP waveguide (**Fig. 8d, e**). In consequence, a heat transfer to the PCM that consists now of the buffer layer is more efficient. Compared to the similar arrangement realized with the photonic waveguide that show higher insertion losses compared to the vertical configuration [42], the losses associated with the presence of the external electrodes in the lateral arrangement are much lower compared to the vertical configuration (**Fig. 7**). For the external electrodes placed only 50 nm from the ridge, the minimum attenuation was calculated at 0.225 dB/µm (**Fig. 7a**), while for a vertical configuration with the heating electrode placed 50 nm below a ridge, the minimum attenuation was calculated at 0.486 dB/µm (**Fig. 7b**). The lateral configuration allows to utilize both electrodes as a heating electrode (**Fig. 7d**) or only one of them (**Fig. 7e**).

*b.   Optical switching and internal heater*

Apart from the external heater(s), the state of PCM that is a part of the waveguide can be reversibly switched between amorphous and crystalline by sending a light (optical pulses) through the LR-DLSPP waveguide (**Fig. 5 and 6**). The guiding properties of the LR-DLSPP waveguide depend on the state of the PCM. Under a design, the LR-DLSPP waveguide can work in a low-loss regime either for PCM in amorphous (**Fig. 7a**, solid blue line) or crystalline state (**Fig. 7b**, solid blue line). For low-loss operation conditions, the LR-DLSPP waveguide is in balance as a mode effective index below a metal stripe (in a ridge) is close to the mode effective index above a metal stripe (in a buffer layer). However, when under either a design or an applied voltage the balance is disturbed, the absorption in metal stripe arises and attenuation increases. Thus, the metal stripe can serve as an internal heater that no require an external voltage (**Fig. 8a**) [50]. In consequence, a heating process can be realized all-optically. Compared to standard on-chip optical switching that cannot switch wide bandgap PCMs that have zero absorption in the near IR (**Fig. 4**) [27], the proposed here arrangement takes an advantage of plasmonics and presence of an internal metal stripe that is part of the waveguide. The optical power absorbed by the metal stripe is dissipated into any materials that are in contact with a metal stripe and an amount of heat dissipated into PCM depends on thermal resistance and capacity of the surroundings materials (**Fig. 8a**). The dissipated power by metal stripe depends on the surface plasmon polariton attenuation coefficient and the length of the active region. Thus, the higher attenuation coefficient, the higher increase of the metal stripe temperature. As it has been previously shown, the temperature increases of the materials that are in contact with a metal stripe is proportional to the power coupled to plasmonic waveguide [50].

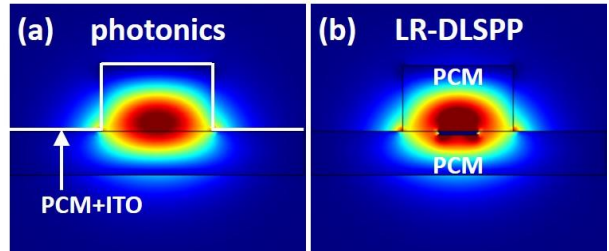

**Figure 9**. Simulated 2D mode profiles of the fundamental TM mode for (a) photonic and (b) LR-DLSPP waveguides with PCM.

Let's now compare a heating mechanism of the photonic waveguide with a proposed LR-DLSPP waveguide in terms of the optical switching.

In a photonic waveguide, a thin layer of a PCM (10 nm GST) is deposited on top of the Si or SiN waveguide and covered with indium tin oxide (ITO) (10 nm thick) to prevent oxidation of the PCM (Fig. 9a) [51, 52]. As observed from **Fig. 9a**, the PCM is placed far away from the electric field maximum of the propagating mode, thus, the interaction of light with a PCM is weak.

In comparison, for a LR-DLSPP waveguide, the PCM can be placed either in a buffer layer or in a ridge that are in direct contact with a metal stripe, *i.e.*, in the electric field maximum of the propagating plasmonic mode (**Fig. 9b**). Furthermore, as the electric field is much more enhanced in the plasmonic waveguide compared to the photonic waveguide, the electric field of the propagating LR-DLSPP mode (**Fig. 9b**) that interact with a PCM is much stronger compared to the photonic waveguide (**Fig. 9a**).

Furthermore, as it was mentioned above, the optical switching in a proposed waveguide can be highly enhanced through a heat transfer from a metal stripe under an absorption of light to the PCM that is in a direct contact with a metal stripe [50]. Thus, two stage heating process of PCM can take place in a proposed LR-DLSPP arrangement that is related with (a) a direct absorption of light by PCM, and (b) a heat transfer from a metal stripe to the PCM under an absorption of light by metal stripe.

*c. Electrical (memory) switching*

Furthermore, when consider either a vertical configuration with the external electrode in direct contact with the PCM placed either in a ridge or a buffer layer (**Fig. 8a**) or lateral with the PCM placed in a buffer layer (**Fig. 8d**), the external electrode and a metal stripe can be in direct contact with a PCM and a distance between electrodes can be below 100 nm. Thus, the PCM is connected into a circuit what enables the electrical (memory) switching. When electrical current flows through PCM that is part of the circuit, the Joule heating of PCM itself causes the phase transition. One of the main limitations of this switching mechanism is that only a very small volume of PCM can be changed as it has been previously show in Ref. 30. In our case, very small volume in the range of hundreds of nanometers is required to attain either full $\pi$ phase delay or provide high optical contrast. Our proposed LR-DLSPP waveguide is very sensitive on any changes of the refractive index of PCM. Under a change of refractive index of PCM that is placed in the ridge from $n = 4.0$ to $n = 5.8$, the attenuation arises from ~0.183 dB/µm to ~2.4 dB/µm, thus only 1 µm long waveguide is required to introduce an optical contrast exceeding 2.2 dB (**Fig. 7b**). Or only 280 nm long waveguide is needed to provide full $\pi$ phase delay with an insertion loss below ~0.12 dB (**Fig. 7b**) when the PCM refractive index changes from $n = 3.2$ and $n = 5$.

The proposed in this paper arrangement provide a nanoscale volume of PCM, which reduces the volume of material that needs to be heated during switching and so in turn reduces switching energies. Furthermore, the process of switching is enhanced by a plasmonic structure. In consequence, the small PCM volume, the high volume/surface ratio, and implementation of plasmonics can contribute to a very effective switching process.

In this work the simulations and calculation were performed for Au as a metal stripe and a heating electrode, however any other "plasmonic" metals [29, 53, 54] can be used as a metal supporting a LR-DLSPP mode and any other metals can be implemented as the external electrodes that can provide even higher benefits in terms of heat generation efficiency.

**Amplitude (absorption) modulation**

The proposed arrangement allows modulation of the optical signal under an applied voltage. The state of PCM that is a part of the waveguide can be reversibly switched between amorphous and crystalline states by sending electrical pulses through the metal stripe and/or metal electrode(s) (**Fig. 5b and 6a**). The guiding properties of the LR-DLSPP waveguide depend on the state of the PCM. For an amorphous state of PCM the LR-DLSPP waveguide works in a low-loss regime as a mode effective index below a metal stripe is close to the mode effective index above a metal stripe. Thus, the LR-DLSPP waveguide is in balance. For a crystalline state of PCM the balance in mode effective indices is broken as one side of the LR-DLSPP waveguide (upper or lower) with PCM shows higher mode effective index. In consequence, the absorption in metal stripe arises resulting in shorter propagation length of the mode and, in turn, lower optical transmission. To take an example, the attenuation change from 0.183 dB/µm to 2.4 dB/µm was calculated under a change of PCM refractive index placed in the ridge from $n = 4.0$ to $n = 5.8$ (**Fig. 7b**, blue solid line). Further enhancement is possible as no optimization was provided at this stage of the research.

Under a change of a refractive index of PCM the mode is pushed to one interface of metal stripe that is characterized by a lower mode effective index. As a distance between a metal stripe and an external electrode can be in a range of from tens to hundreds of nanometers, the mode propagating on one side of a metal stripe can couple to the external electrode, thus creating a gap surface plasmon polariton mode propagating in the metal-PCM-metal structure. Compared to the LR-DLSPP mode that is characterized by a long propagation distance in the range of hundreds of micrometers [31, 32, 33, 34], the MIM mode propagation distance usually no exceed ten of micrometers [14]. Thus, under a phase transition of the PCM, a huge difference in a propagation length of the plasmonic mode can be observed ranging from a few micrometers to up to hundreds of micrometers. In comparison, a previously reported MIM plasmonic modulation with the GST placed in the gap showed the optical contrast (a change in a light transmission) lower than 1 % at telecom wavelength of 1550 nm under a transition of GST from an amorphous to a crystalline state (**Table 2**) [30]. As observed on this example and Table 2, the absorption modulation of the PCM-based photonic straight waveguides is not very impressive, thus, PCMs are often integrated with photonic cavities where both optical phase and absorption modulation effect of the PCMs contribute to high extinction ration [36].

**Table 2.** Comparison of different absorption modulation platforms with PCMs on SOI at 1550 nm. Here, $L$ is the total length of the amplitude modulator, $IL$ is the insertion loss, $ER$ is the extinction ratio, and distance to heater defines a distance between external heater and ridge.

|  | $L$ (µm) | $IL$ (dB) | $ER$ [dB] | Distance to heater [nm] | Nonvolatility/dual-mode operation? |
|---|---|---|---|---|---|
| **GST with PIN diode heater [48]** | 4 | 0.037 | 5 | 500 | Yes/No |
| **20 nm SbS with ITO heater [49]** | 10 | 0.087 | 1.77 | <500 | Yes/No |
| **MZI with GSST on Si waveguide [42]** | 3.84 | 3 | ~8 | 250 | Yes/No |
| **Plasmonic nanogap with GST [30]** | ~0.1 | 9 | ~1 % | 50 | Yes/Yes |
| **LR-DLSPP with zero-loss PCMs (this work)** | **3** | **0.55** | **6.65** | **90** | **Yes/Yes** |

Here it should be emphasized that results provided for the LR-DLSPP waveguide in an absorption modulation schema are based on an assumption of zero loss PCMs for both an amorphous and crystalline states. Thus, the results mostly refer here to low-loss or even zero-loss PCMs such as, for example, $Sb_2S_3$ and $Sb_2Se_3$. If GST or GSST integrated in such a LR-DLSPP waveguide, a huge enhancement in the extinction ration should be observed.

The proposed modulation arrangement that is based on the PCM can minimize the switching voltage and current as the PCM is placed in the electric field maximum of the propagating mode. Thus, it provides a strong light confinement inside the PCM without introducing excessive optical losses (**Table 2**) [40]. Furthermore, as the PCM is in direct contact with the electrodes it enhances heat localization within the PCM to effectively lower thermal mass and it enhances the speed of the switching the PCM device [40].

**Phase modulation**

The main obstacles in achieving phase modulation is due to the small refractive index perturbation what result in a large device footprint, significant optical losses and long switching time. Moreover, all the switching mechanism are volatile what require a constant power supply to operate [10, 15, 16, 17, 18, 19, 20].

Phase modulation require interferometric schema such as for example Mach Zehnder Interferometers [12, 13, 17, 19]. The operation of the MZI is based on changing the mode propagation constant in one arm resulting in the phase difference of two modes interfering at the output Y-junction/ the length of the MZI arm required to ensure complete modulation, *i.e.*, switching the light off in the Y-junction, is related to the phase difference π between the arms (**Fig. 1**).

Low-loss PCMs such as for example $Sb_2S_3$ and $Sb_2Se_3$ (**Fig. 4b**) enable a realization of nonvolatile phase modulation with LR-DLSPP waveguides where the change of phase of PCM effect rather the mode effective index than attenuation of the LR-DLSPP waveguide (**Fig. 7a**, green dashed line). In this example, the attenuation for both PCM phases is kept constant at 0.28 dB/μm, however further reduction is possible as no optimization was provided at this stage of the research. While the attenuation is kept constant on the same level, the mode effective index change $\Delta n_{eff}$ = ~0.83 was calculated that corresponds to a phase shift of ~1.07 π/μm. Thus, only ~930 nm in-length phase shifter is required to attain full π phase delay. The small device footprint also yields a low theoretical insertion loss of ~0.26 dB when switching the refractive index of PCM from $n$ = 3.47 to $n$ = 5.25.

**Table 3.** Comparison of different phase shifter platforms (PCM-based and others) on SOI at 1550 nm. Here, $\Delta n_{eff}$ refers to the change in effective index, $L_\pi$ is the total length needed to achieve a π phase shift, *IL* is the insertion loss, $\alpha_{att}$ refers to the attenuation under a phase shift, distance to heater defines a distance between external heater and ridge.

| | $\Delta n_{eff}$ | $L_\pi$ (μm) | IL (dB) | $\alpha_{att}$ [dB] | Distance to heater [nm] | Nonvolatility/dual-mode operation? |
|---|---|---|---|---|---|---|
| **Thermo-optical with doped Si [55]** | $1.5 \times 10^{-4} \times \Delta T$ | 61.6 | 0.23 | - | - | No /No |
| **Electro-optical with injection [56]** | ~$10^{-3}$ | 400 | 1.5 | - | - | No/No |
| **Optoelectro-mechanical [57]** | ~0.1 | 210 | 0.47 | - | - | No/No |
| **BTO Pockels effect [58]** | $7 \times 10^{-4}$ | 1000 | 1 | - | - | No/No |
| **Plasmonic with nonlinear polymer [59]** | 0.055 | 29 | 12 | - | - | No/No |
| **Lithium niobate [60]** | $1.5 \times 10^{-4}$ | 5000 | 2.5 | - | - | No/No |
| **20 nm SbS with ITO heater [49]** | 0.02 | 38.7 | 0.34 | 6.8 | <500 | Yes/No |
| GST with PIN diode heater [48] | 0.08 | 9.7 | 0.36 | 16 | 500 | Yes/No |
| 30 nm GSST with graphene heater [23] | 0.28 | 2.77 | 14.2 | - | <500 | Yes/No |
| GST with graphene heater [24] | 0.21 | 3.6 | 0.5 | >22 | 500 | Yes/No |
| 30 nm SbS with Si-doped heater [25] | 0.071 | 11 | 0.45 | 0.07 | 500 | Yes/No |
| **LR-DLSPP with PCM (vertical) (this work)** | **2.75** | **0.28** | **0.12** | **0** | **50** | **Yes/Yes** |
| **LR-DLSPP with PCM (lateral) (this work)** | **0.83** | **0.93** | **0.26** | **0** | **>150** | **Yes/Yes** |

In **Table 3** we compare the proposed here PCM-based plasmonic phase shifter with the state-of-the-art PCM phase shifters and other phase shifters realized in different technologies in terms of the change in effective index, $\Delta n_{eff}$, the total length to achieve a $\pi$ phase shift, $L_\pi$, the insertion losses, $IL$, the accompanying attenuation, $\alpha_{att}$, and a distance from a heater to the waveguide. In terms of the PCM phase shifters, a distance between a heater and waveguide defines the energy/power consumption and the switching time as less generated heat is dissipated around the long this distance. Thus, the smaller distance the lower energy/power required to achieve a $\pi$ phase shift and faster switching time. On the other hand, most of a traditional photonic PCM phase shifters place the electrodes far away from the waveguide to avoid high insertion losses [23, 24, 25, 48, 49]. Thus, a trade-off between the insertion loss and the energy efficiency exists. In the case of the proposed here plasmonic PCM phase shifter this trade-off can be largely avoided as a plasmonic mode is tightly bounded to the metal stripe, thus, the external electrode(s) can be placed very close to the plasmonic waveguide (see **Table 3**).

The static phase-shifter or phase modulator that does not consume any energy to hold the state is a key component of the on-chip self-reconfigurable optical network that can perform any linear operation or couple between any input and output. Most of the available or proposed switches suffer from high insertion loss, slow switching speed, small modulation depth, and high energy consumption (**Table 3**). Thus, the proposed here a phase shifter operating in a low loss regime and under a low power consumption is under a deep interest.

**Losses of the devices**
For an optimum operation conditions, the overall losses of the system need to be kept at the minimum. The figure of merit (FoM) in Mach-Zehnder Modulators (MZM) that are based on the Mach-Zehnder Interferometer (MZI) arrangement is the product of the half-wave voltage and the active modulator length, $V_\pi L$ [17, 19]. However, as some of the plasmonic modulators show very good FoM, most of them, simultaneously, suffer from very high insertion and coupling losses to the active modulator area exceeding 8 dB [16, 17, 22, 30]. Thus, even being very compact, the insertion losses exceeding 4.5 dB for just 3 μm long active region were measured for ITO deposited on top of the waveguide [17]. Simultaneously, a doped silicon heater with PCM lead to high insertion losses exceeding ~0.5 dB/μm [43]. Therefore, most of them become impractical for large-scale PIC platforms where light is guided through numerous photonic routers.

The proposed modulators can provide extremally low insertion losses even below 0.0075 dB/μm when the external electrodes are placed 500 nm away from the ridge (**Fig. 3a**). As we can expect that some imperfections in the fabrication process can increase this value to about 0.02 dB/μm, it still exceeds previously reported values for other plasmonic modulators [16, 17, 18, 22]. Thus, even assuming a 5 μm long active area of the phase modulator, the insertion losses below 0.1 dB can be obtained. Simultaneously, the coupling losses between photonic waveguide and LR-DLSPP waveguide per interface as low as ~0.05 dB were calculated [31, 32]. Assuming again some imperfections in the fabrication process, the value ~0.2 dB per interface seems reasonable. In conclusion, even 5 μm long photodetector can provide the overall losses below 0.5 dB under an assumption of some imperfections during a fabrication process. However, from a theoretical point of view, the overall losses as low as ~0.115 dB can be achieved.

As it has been previously shown the proposed arrangement can be used as well as a photodetector and an activation function unit [61] or/and can serve as a building block for future plasmonic neural networks [62].

**Energy consumption**

As the energy-per-bit scales with the length of the modulator and power consumption required to switch its state, the proposed arrangement that is based on the LR-DLSPP waveguide offers high reduction in energy consumption. Direct contact of the external or internal heater(s) with a PCM that has very strong influence on the propagating mode, highly reduces power requirements and enables low voltage operation (**Table 2 and 3**). As it was shown in this paper, the length of the proposed switches operating base on the amplitude or phase modulation effects do not exceed 1 μm.

**Conclusion**

In conclusion, we have proposed a new class of electrically- and, preferably, optically-driven plasmonic nonvolatile switches operating based on a phase shifts and/or amplitude modulation that achieve zero-static power consumption in a device being extremally compact. Such a nonvolatile phase shifter requires only ~230 nm long active area to attain full $\pi$ phase delay while an insertion loss is kept below ~0.12 dB. In comparison, a nonvolatile switch operating on an amplitude modulation can show an extinction ratio exceeding ~2.2 dB/μm and insertion loss of ~0.185 dB/μm. Compared to any other nonvolatile devices that are based on the phase change materials, the heating of material can be realized through either external or internal heater(s), memory switching or optical switching. They can be exploited separately or linked to provided further enhancement of temperature increases.

**Author information**
**Affiliations**
Independent Researcher, 90-132 Lodz, Poland
Jacek Gosciniak
**Contributions**
J.G. conceived the idea, performed all calculations, FEM and FDTD simulations and wrote the article.
**Corresponding author**
Correspondence to Jacek Gosciniak (jeckug10@yahoo.com.sg)